\documentclass[graybox]{svmult}

\usepackage{amsmath, amssymb}
\usepackage{mathptmx}       % selects Times Roman as basic font
\usepackage{helvet}         % selects Helvetica as sans-serif font
\usepackage{courier}        % selects Courier as typewriter font
\usepackage{type1cm}        % activate if the above 3 fonts are
\usepackage{}
\usepackage{makeidx}         % allows index generation
\usepackage{graphicx}        % standard LaTeX graphics tool
\usepackage{multicol}        % used for the two-column index
\usepackage[bottom]{footmisc}% places footnotes at page bottom

\newcommand{\R}{{\mathbb R}}
\newcommand{\N}{{\mathbb N}}
\newcommand{\Z}{{\mathbb Z}}
\newcommand{\C}{{\mathbb C}}

\newcommand{\sgn}{\mathop{\rm sgn}}
\newcommand{\A}{{\mathbf A}}

\renewcommand{\geq }{\geqslant}

\renewcommand{\leq }{\leqslant}

\makeindex             % used for the subject index
\begin{document}

\title*{Spherical Schr\"odinger Hamiltonians: Spectral Analysis and Time Decay}
\author{Luca Fanelli}
\institute{Luca Fanelli \at Name, Dipartimento di Matematica, SAPIENZA Universit\`a di Roma,
P.~le Aldo Moro 5, 00185 Roma, \email{fanelli@mat.uniroma1.it}}
\maketitle

%\abstract*{This is a survey of recent results concerning with the description of the canonical dispersive flow $e^{itH}$ led by Schr\"odinger Hamiltonian $H$. We study, in particular, how the time decay of space $L^p$-norms depends on the frequency localization of the initial datum. A quite complete description of the phenomenon is given in terms of the spectral features of the restriction of $H$ to the unit sphere, and an analogy with some uncertainty inequality is presented.}

\abstract{In this survey, we review recent results concerning the canonical dispersive flow $e^{itH}$ led by a Schr\"odinger Hamiltonian $H$. We study, in particular, how the time decay of space $L^p$-norms depends on the frequency localization of the initial datum with respect to the some suitable spherical expansion. A quite complete description of the phenomenon is given in terms of the eigenvalues and eigenfunctions of the restriction of $H$ to the unit sphere, and a comparison with some uncertainty inequality is presented.}

\section{Introduction}
\label{sec:1}
For $\psi=\psi(t,x):\R\times\R^d\to\C$, 
let us consider the free Schr\"odinger equation
\begin{equation}\label{eq:schro}
\partial_t\psi=i\Delta \psi,
\qquad
\psi(0,x)=\psi_0(x).
\end{equation}
Solving \eqref{eq:schro} with initial datum $\psi_0(x)\in L^2(\R^d)$ is to find a wavefunction $\psi\in\mathcal C^1(\R;L^2(\R^d))$ such that $\widehat\psi(t,\xi)=e^{-it|\xi|^2}\widehat\psi_0(\xi)$, the hat denoting the Fourier transform in the $x$-variable
$$
\widehat\psi(t,\xi):=\int_{\R^d}e^{-it\xi}\psi(x)\,dx.
$$
Computing the distributional Fourier transform of $e^{-it|\xi|^2}$, one finds that the unique solution to \eqref{eq:schro}, in the above sense, is given by
\begin{equation}\label{eq:fundsol}
\psi(t,x)=(4\pi it)^{-\frac d2}e^{-i\frac{|x|^2}{4t}}*\psi_0(x)
=(4\pi it)^{-\frac d2}e^{-i\frac{|x|^2}{4t}}\int_{\R^d}e^{i\frac{x\cdot y}{2t}}e^{-i\frac{|y|^2}{4t}}\psi_0(y)\,dy.
\end{equation}
From now on, we will denote by $e^{it\Delta}$ the one-parameter flow on $L^2(\R^d)$ defined by formula \eqref{eq:fundsol}, namely $e^{it\Delta}\psi_0(\cdot)=\psi(t,\cdot)$, being $\psi$ as in \eqref{eq:fundsol}. By Plancherel Theorem it follows that $e^{it\Delta}$ is unitary on $L^2(\R^d)$, namely
\begin{equation}\label{eq:cons}
\left\|e^{it\Delta}\psi_0(\cdot)\right\|_{L^2(\R^d)}=\|\psi_0\|_{L^2(\R^d)},
\qquad
\forall t\in\R.
\end{equation}
By \eqref{eq:fundsol}, it also immediately follows that
\begin{equation}\label{eq:decay}
\left\|e^{it\Delta}\psi_0(\cdot)\right\|_{L^\infty(\R^d)}\leq C|t|^{-\frac d2}\|\psi_0\|_{L^1(\R^d)},
\qquad
\forall t\in\R,
\end{equation}
with a constant $C>0$ independent on $t$ and $\psi_0$. The last inequality, together with \eqref{eq:cons}, gives by Riesz-Thorin the full list of {\it time decay estimates} for the free Schr\"odinger equation
\begin{equation}\label{eq:decayfull}
\left\|e^{it\Delta}\psi_0(\cdot)\right\|_{L^p(\R^d)}\leq C|t|^{-d\left(\frac12-\frac1p\right)}\|\psi_0\|_{L^{p'}(\R^d)},
\qquad
\forall t\in\R,
\qquad
\forall p\geq2
\end{equation}
where the constant $C$ only depends on $p$ and $d$. Inequalities \eqref{eq:decayfull} turn out to be a crucial tool in Scattering Theory and Nonlinear Analysis; in particular, a suitable time average of the same leads to the so called {\it Strichartz estimates} (see the standard reference \cite{KT}), which play a fundamental role both for fixed point results and as Restriction Theorems for the Fourier transform:
\begin{equation}\label{eq:stri}
\left\|e^{it\Delta}\psi_0\right\|_{L^q_tL^r_x}\leq C\|\psi_0\|_{L^2(\R^d)},
\end{equation}
with $2/q=d/2-d/r$, $q\geq2$ and $(q,r,d)\neq(2,\infty,2)$, and
\begin{equation*}
\left\|e^{it\Delta}\psi_0(\cdot)\right\|_{L^p(\R^d)}
:=
\left\|\left\|e^{it\Delta}\psi_0(\cdot)\right\|_{L^r(\R^d)}\right\|_{L^q(\R)}.
\end{equation*}
From now on, we point our attention on estimate \eqref{eq:decay} and try to give it a deeper insight. First of all, it is clear by \eqref{eq:fundsol} that a crucial role is played by the plane wave $K(x,y):=e^{i\frac{x\cdot y}{2t}}$ which is uniformly bounded with respect to the $x,y$ variables, for any fixed time $t\neq0$, i.e.
\begin{equation}\label{eq:condiz}
\sup_{x,y\in\R^d}\left|e^{i\frac{x\cdot y}{2t}}\right|=1<\infty,
\qquad
\forall t\neq0.
\end{equation}
We stress that a completely analogous behavior occurs when one solves, for positive times, the Heat Equation
\begin{equation}\label{eq:heat}
\partial_t u = \Delta u,
\qquad
u(0,x) = u_0(x)\in L^p(\R^d),
\end{equation}
since the solution is given by the convolution 
\begin{equation}\label{eq:fundheat}
u(t,x)=(4\pi t)^{-\frac d2}e^{\frac{-|x|^2}{4t}}* u_0(x),
\qquad
(t>0)
\end{equation}
for all $p\in[1,+\infty]$. This shows that \eqref{eq:heat} satisfies the same a priori estimates \eqref{eq:decayfull} as equation \eqref{eq:schro}. Notice that \eqref{eq:schro} and \eqref{eq:heat} enjoy the same scaling invariance
$$
f_\lambda(t,x):=f\left(\frac{t}{\lambda^2},\frac{x}{\lambda}\right),
\qquad
\lambda>0.
$$
In addition, the Gaussian decay in \eqref{eq:fundheat} is much smoother than the oscillating character of the fundamental solution in \eqref{eq:fundsol}, and leads to much stronger phenomena than the ones led by the dispersive flow $e^{it\Delta}$. Nevertheless, from the point of view of estimate \eqref{eq:decay} the behavior is the same for the flows $e^{t\Delta}, e^{it\Delta}$, when $t>0$.
Our first question is the following: 
\begin{itemize}
\item[{\bf A}]
{\it is the time decay of the flows $e^{t\Delta},e^{it\Delta}$ related to the lowest frequency behavior of the corresponding fundamental solutions?}
\end{itemize}
We now pass to a more precise analysis of the decay estimate in \eqref{eq:decay}, to describe some additional phenomenon which is hidden in formula \eqref{eq:fundsol}. To this aim, let us recall the {\it Jacobi-Anger} expansion or plane waves, which combined with the Addition
Theorem for spherical harmonics (see for example \cite[formula (4.8.3), p.
116]{Ismail} and \cite[Corollary 1]{BeStr}) yields
\begin{equation}\label{eq:jacobi}
e^{ix\cdot y}=(2\pi )^{d/2}\big(|x||y|\big)^{-\frac{d-2}{2}}\sum_{\ell
=0}^{\infty }i^{\ell }J_{\ell +\frac{d-2}{2}}\big(|x||y|\big)\bigg(%
\sum_{m=1}^{m_{\ell }}Y_{\ell ,m}\big(\tfrac{x}{|x|}\big)\overline{%
Y_{\ell ,m}\big(\tfrac{y}{|y|}\big)}\bigg)
\end{equation}%
for all $x,y\in {\mathbb{R}}^{d}$. Here $J_\nu$ denotes the $\nu-$th Bessel function of the first kind
\begin{equation*}
J_{\nu }(t)=\bigg(\frac{t}{2}\bigg)^{\!\!\nu }\sum\limits_{k=0}^{\infty }
\dfrac{(-1)^{k}}{\Gamma (k+1)\Gamma (k+\nu +1)}\bigg(\frac{t}{2}\bigg)
^{\!\!2k}
\end{equation*}
and the $Y_{\ell,m}$ are usual spherical harmonics. Recalling that $J_{\nu}(t)\sim t^\nu$, for $\nu\geq0$, as $t$ goes to 0, we see that an additional time-decay, for $t$ large is hidden in formula \eqref{eq:fundsol}, in the term $e^{i\frac{x\cdot y}{t}}$. Roughly speaking, we expect that initial data which are localized higher frequencies (with respect to the spherical harmonics expansion) decay polynomially faster along a Schr\"odinger evolution, in suitable topologies. This leads to our second question:
\begin{itemize}
\item[{\bf B}]
{\it how can the above described phenomenon be quantified, and how stable is it under lower-order perturbations?}
\end{itemize}
Looking to identity \eqref{eq:jacobi}, the presence of spherical harmonics and special functions gives the hint that the spherical laplacian is playing an important role in the description of the above mentioned phenomena. 
The aim of this survey is to describe this role, giving partial answers to the above questions and leaving some open problems, corroborated by some recent results.

\section{A stationary viewpoint: Hardy's Inequalitiy}
\label{sec:2}
We devote a preliminary section to introduce an interesting stationary viewpoint of the above picture, related to some uncertainty inequalities. To this aim, we recall the well known {\it Hardy's inequality}:
\begin{equation}\label{eq:hardy}
\frac{(d-2)^2}{4}\int_{\mathbb R^d}\frac{|\psi(x)|^2}{|x|^2}\,dx
\leq
\int_{\mathbb R^d}|\nabla\psi(x)|^2\,dx,
\qquad
(d\geq3)
\end{equation}
which holds for any function $\psi$ such that $|\nabla\psi|\in L^2$.
The constant in front of inequality \eqref{eq:hardy} is sharp, and it is not attained on any function $\psi$ for which the right-hand side is finite, as we see in a while. Inequality \eqref{eq:hardy} can be rewritten in operator terms as
\begin{equation}\label{eq:hardyop}
-\Delta -\frac{\lambda}{|x|^2}\geq0,
\qquad
\forall \lambda\leq\frac{(d-2)^2}{4}
\qquad
(d\geq3).
\end{equation}
This has to be interpreted in the sense of the associated quadratic form.
The proof of \eqref{eq:hardy} relies on the following fact: given a symmetric operator $\mathcal S$ and a skew-symmetric operator $\mathcal A$ on $L^2$, one can (formally) compute
\begin{equation*}
0\leq\int_{\mathbb R^d}\left|(\mathcal A+\mathcal S)\psi\right|^2\,dx
=\int_{\mathbb R^d}|\mathcal A\psi|^2\,dx+\int_{\mathbb R^d}|\mathcal S\psi|^2\,dx
-\int_{\mathbb R^d}\overline\psi\left[\mathcal A,\mathcal S\right]\psi\,dx,
\end{equation*}
where $[\mathcal A,\mathcal S]=\mathcal A\mathcal S-\mathcal S\mathcal A$.
Then the choices
\begin{equation*}
\mathcal A:=\nabla,
\qquad
\mathcal S:=\frac{d-2}{2}\frac{x}{|x|^2}
\qquad
\Rightarrow
\qquad
[\mathcal A,\mathcal S]=\frac{(d-2)^2}{2|x|^2}
\end{equation*}
immediately give \eqref{eq:hardy} for functions $\psi$ smooth enough, and a regularization argument completes the proof. Also notice the equality in \eqref{eq:hardy} is attained when $(\mathcal A+\mathcal S)\psi\equiv0$, which yields the maximizing function $\psi(x)=|x|^{1-\frac d2}$, and we see that $|\nabla\psi|\notin L^2$, as mentioned above. In addition, one immediately realizes that, given $\widetilde{\mathcal A}=\partial_r=\nabla\cdot\frac{x}{|x|}$, then
\begin{equation*}
[\widetilde{\mathcal A},\mathcal S]=[\mathcal A,\mathcal S]=\frac{(d-2)^2}{2|x|^2},
\end{equation*}
which yields the more precise inequality
\begin{equation}\label{eq:hardyrad}
\frac{(d-2)^2}{4}\int_{\mathbb R^d}\frac{|\psi(x)|^2}{|x|^2}\,dx
\leq
\int_{\mathbb R^d}|\partial_r\psi(x)|^2\,dx,
\qquad
(d\geq3)
\qquad
\end{equation}
In other words, inequality \eqref{eq:hardyrad} shows that the angular component of $-\Delta$ is not playing a role in \eqref{eq:hardy}-\eqref{eq:hardyop}. To understand this fact, it is convenient to use spherical coordinates and write
\begin{equation}\label{eq:spher}
\Delta=\partial_r^2+\frac{d-1}{r}\partial_r+\frac{1}{r^2}\Delta_{\mathbb S^{d-1}},
\end{equation}
being $\Delta_{\mathbb S^{d-1}}$ the spherical laplacian, i.e. the Laplace-Beltrami operator on the $(d-1)$-dimensional unit sphere. We recall that $-\Delta_{\mathbb S^{d-1}}$ is a (positive) operator with compact inverse, hence it has purely point spectrum which accumulates at infinity, which is explicitly given by the set
\begin{equation}\label{eq:specfree}
\sigma\left(-\Delta_{\mathbb S^{d-1}}\right)=\sigma_{\textrm{p}}\left(-\Delta_{\mathbb S^{d-1}}\right)
=\{\ell(\ell+d-2)\}_{\ell=0,1,2,\dots}.
\end{equation}
Spherical harmonics $\{Y_{\ell,m}\}$ are associated eigenfunctions, which form a complete orthonormal set in $L^2(\mathbb S^{d-1})$. Denoting by $H_\ell$ the eigenspace associated to the $\ell$-th eigenvalue of $-\Delta_{\mathbb S^{d-1}}$, by $D_\ell$ its algebraic dimension, and by $H_{\ell,m}$ the space generated by $Y_{\ell,m}$, we have the well known decomposition
\begin{equation*}
L^2(\mathbb S^{d-1})=\bigoplus_{\stackrel{l\geq0}{1\leq m\leq D_\ell}}H_{\ell,m}
\end{equation*}
Therefore any function $\psi\in L^2(\mathbb R^d)$ has a (unique) expansion 
\begin{equation}\label{eq:expansion}
  \psi(x)=\sum_{\ell=0}^{\infty}\sum_{m=1}^{D_\ell}\psi_{\ell,m}(r)Y_{\ell,m}(\omega)
  \qquad
  x=r\omega,\ \ \ r:=|x|
\end{equation}
and moreover
\begin{equation*}
\|f(r\omega)\|_{L^2(\mathbb S^{d-1})}=\sum_{\stackrel{\ell\geq0}{1\leq m\leq D_\ell}}|f_{\ell,m}|^2.
\end{equation*}

We can hence use \eqref{eq:spher} to write
\begin{align}\label{eq:hardyspher}
&
\int_{\mathbb R^d}|\nabla\psi|^2\,dx
=-\int_{\mathbb R^d}\overline\psi\Delta\psi\,dx
\\
&\ \ \ \ 
=\underbrace{-\int_{\mathbb R^d}\overline\psi\left(\partial_r^2\psi+\frac{d-1}{r}\partial_r\psi\right)\,dx}_{=:I}+\underbrace{\int_{\mathbb R^d}\frac{1}{|x|^2}\left\langle\psi,-\Delta_{\mathbb S^{d-1}}\psi\right\rangle_{L^2(\mathbb S^{d-1})}\,dx.}_{=:II}
\nonumber
\end{align}
where the brackets $\langle\cdot,\cdot\rangle_{L^2(\mathbb S^{d-1})}$ denote the inner product in $L^2(\mathbb S^{d-1})$. Arguing as above we see that
\begin{equation*}
I\geq\frac{(d-2)^2}{4}\int_{\mathbb R^d}\frac{|\psi(x)|^2}{|x|^2}\,dx,
\qquad
(d\geq3)
\end{equation*}
which is inequality \eqref{eq:hardyrad}. On the other hand, it follows by \eqref{eq:specfree} that
\begin{equation*}
II\geq0,
\end{equation*}
therefore no additional contribution to \eqref{eq:hardy} is given by $-\Delta_{\mathbb S^{d-1}}$. Nevertheless, given $\psi\in L^2(\mathbb R^{d-1})$, if $\psi_{0,1}=0$ in the expansion \eqref{eq:expansion} (notice that $H_{0,1}$ coincides with the space of $L^2$-radial functions), then by \eqref{eq:specfree} it follows that
\begin{equation*}
\left\langle\psi,-\Delta_{\mathbb S^{d-1}}\psi\right\rangle_{L^2(\mathbb S^{d-1})}
\geq
(d-1)\|\psi(\omega)\|_{L^2(\mathbb S^{d-1})}
\qquad
\text{if }\psi_{0,1}=0
\end{equation*}
and inequality \eqref{eq:hardyrad} improves: 
\begin{equation}\label{eq:hardyrad2}
\int_{\mathbb R^d}|\partial_r\psi(x)|^2\,dx
\geq
\left(\frac{(d-2)^2}{4}+(d-1)\right)\int_{\mathbb R^d}\frac{|\psi(x)|^2}{|x|^2}\,dx,
\qquad
(d\geq2)
\qquad
\psi_{0,1}=0.
\end{equation}
Notice that the previous gives a non trivial 2D-inequality, holding on functions $\psi$ which are orthogonal to $L^2$-radial functions.
More in general, given $\psi\in L^2(\mathbb R^{d})$, let 
\begin{equation*}
\ell_0:=\min\{\ell\in\mathbb N \text{\ such that }\exists m=1,\dots,D_\ell\ :\ \psi_{\ell,m}\neq0\}.
\end{equation*}
Then, by \eqref{eq:hardyspher}, the following Hardy's inequality holds:
\begin{equation}\label{eq:hardyrad3}
\int_{\mathbb R^d}|\partial_r\psi(x)|^2\,dx
\geq
\left(\frac{(d-2)^2}{4}+\ell_0(\ell_0+d-2)\right)\int_{\mathbb R^d}\frac{|\psi(x)|^2}{|x|^2}\,dx.
\qquad
(d\geq1)
\end{equation}
Inequality \eqref{eq:hardyrad3} is a quantitative stationary manifestation of the phenomenon described by question {\bf B} in the Introduction. Here it is clear that the improvement comes from the angular component of the free Hamiltonian. In addition, the above arguments clearly suggest that the sharp constant in front of inequality \eqref{eq:hardyrad3} only depends the {\it lowest energies}, which is reminiscent of question {\bf A} in the Introduction. 

Having this in mind, we now see how linear lower-order perturbations of the free spherical Hamiltonian can perturb the spectral picture in \eqref{eq:specfree}, with consequences on the Hardy's inequality \eqref{eq:hardyrad3}.

\begin{example}[{\bf 0-order perturbations}] For $a\in\mathbb R$, consider the shifted Hamiltonians in dimension $d\geq3$
 \begin{equation*}
   H=-\Delta+\frac{a}{|x|^2},
   \qquad
   L=-\Delta_{\mathbb S^{d-1}}+a.
 \end{equation*}
 Clearly $L$ only has point spectrum, which is just a shift of \eqref{eq:specfree}
 \begin{equation*}
\sigma\left(L\right)=\sigma_{\textrm{p}}\left(L\right)
=\{\ell(\ell+d-2)+a\}_{\ell=0,1,2,\dots}
\end{equation*}
and spherical harmonics are still eigenfunctions. The corresponding Hardy's inequality is trivially
\begin{equation}\label{eq:trivial}
\left(\frac{(d-2)^2}{4}+a\right)\int_{\mathbb R^d}\frac{|\psi(x)|^2}{|x|^2}\,dx
\leq
\int_{\mathbb R^d}|\nabla\psi(x)|^2\,dx
+
a\int_{\mathbb R^d}\frac{|\psi(x)|^2}{|x|^2}\,dx.
\qquad
(d\geq3)
\end{equation}
More in general, if $a=a(\omega):\mathbb S^{d-1}\to\mathbb R$, then it is still true that $L$ as only point spectrum, but the picture is more complicated. A typical phenomenon is the formation of clusters of eigenvalues around the (shifted) free eigenvalues. The size of the clusters depends on some universal dimensional quantity related to $a(\omega)$ (see e.g. the standard references \cite{B,Gu, TV, TW, Weinstein} and Lemma \ref{lem:spectral} below). Moreover, for the lowest eigenvalue of $L$ we have
\begin{equation*}
\mu_0:=\min\sigma\left(L\right)=\inf_{\omega\in\mathbb S^{d-1}}\alpha(\omega).
\end{equation*}
 One easily see by the same arguments as above that the following Hardy's inequality holds
\begin{equation}\label{eq:trivial2}
\left(\frac{(d-2)^2}{4}+\mu_0\right)\int_{\mathbb R^d}\frac{|\psi(x)|^2}{|x|^2}\,dx
\leq
\int_{\mathbb R^d}\overline\psi H\psi\,dx.
\end{equation}
 \end{example}
\begin{example}[{\bf $1^{\text{st}}$-order perturbations}] 
Let $A\in L^2_{\text{loc}}(\R^d)$, and recall the {\it diamagnetic inequality}
\begin{equation*}
|(-i\nabla+A)\psi(x)|\geq|\nabla|\psi(x)||.
\end{equation*}
This
gives for free, together with \eqref{eq:hardy}, that
\begin{equation}\label{eq:hardymagn}
\frac{(d-2)^2}{4}\int_{\mathbb R^d}\frac{|\psi(x)|^2}{|x|^2}\,dx
\leq
\int_{\mathbb R^d}|(-i\nabla+A)\psi(x)|\,dx,
\qquad
(d\geq3).
\end{equation}
We wonder if an improvement to the best constant of inequality \eqref{eq:hardymagn} can occur, due to the presence of an angular perturbation of the associated Hamiltonian, in the same style as in the above example.
The main example we have in mind is given by the 2D-{\it Aharonov-Bohm} vector potential:
for $\lambda\in\mathbb R$, consider let us denote by
\begin{equation*}
A:\mathbb R^2\to\mathbb R^2,
\qquad
A(x,y):=\lambda\left(\frac{x}{x^2+y^2},\frac{y}{x^2+y^2}\right)
\end{equation*}
and consider the following quadratic form
\begin{equation*}
q[\psi]:=\int_{\mathbb R^2}|(-i\nabla+A)\psi|^2\,dx.
\end{equation*}
Since $q$ is positive, we can consider the {\it Friedrichs' extension} of the self-adjoint Hamiltonian $H:=-\nabla_A^2$, on the natural form domain induced by $q$ (see Section \ref{sec:main} below for details). The angular component of $H$ is the operator
\begin{equation*}
L:=\left(-i\nabla_{\mathbb S^{1}}+\mathcal A(\omega)\right)^2,
\qquad
\mathcal A:\mathbb S^1\to\mathbb S^1,
\qquad
\mathcal A(x,y)=\lambda\left(\frac{x}{\sqrt{x^2+y^2}},\frac{y}{\sqrt{x^2+y^2}}\right).
\end{equation*}
As above, $L$ has compact inverse and its spectrum is explictly given by
\begin{equation*}
\sigma(L)=\sigma_{\textrm{p}}(L)=\{(\lambda-z)^2\}_{z\in\mathbb Z}.
\end{equation*}
Therefore, the lowest eigenvalue is given by
\begin{equation*}
\mu_0:=\min\sigma(L)=\text{dist}\,(\lambda,\mathbb Z)^2\geq0
\end{equation*}
and we gain the following 2D-Hardy's inequality, proved in \cite{lw}
\begin{equation}\label{eq:hardylw}
\mu_0\int_{\mathbb R^2}\frac{|\psi(x)|^2}{|x|^2}\,dx
\leq
\int_{\mathbb R^2}|(-i\nabla+A)\psi|^2\,dx.
\end{equation}
As soon as $\lambda\notin \mathbb Z$, this is an improvement with respect to the free case $A\equiv0$, in which such an inequality cannot hold for any function $\psi$ such that $|\nabla\psi|\in L^2(\R^2)$ (since the weight $|x|^{-2}$ is not locally integrable in 2D).
\end{example}
In view of the above considerations, we will restrict our attention, from now on, to some scaling-critical electromagnetic Hamiltonians and we will present some recent results which partially answer to questions {\bf A} and {\bf B} in the Introduction of this survey.

\section{Decay estimates: main results}\label{sec:main}

From now on, for any $x\in\mathbb R^d$, we denote by $x=r\omega$, $r=|x|$.
Let 
\begin{equation*}
  \mathbf{A}=\mathbf A(\omega):\mathbb S^{d-1}\to\mathbb R^d,
  \qquad
  a=a(\omega):\mathbb S^{d-1}\to\mathbb R
\end{equation*}
to 0-degree homogenqous functions, and consider the quadratic form 
\begin{equation}\label{eq:quaform}
q[\psi]:=\int_{\mathbb R^d}\left|\left(-i\nabla+\frac{A(\omega)}{r}\right)\psi(x)\right|^2\,dx
+\int_{\mathbb R^d}\frac{a(\omega)}{r^2}|\psi(x)|^2\,dx.
\end{equation}
As we see in the sequel, under suitable conditions, a self-adjoint Hamiltonian 
\begin{equation}\label{eq:hamiltonianH}
H:=\left(-i\nabla+\frac{A(\omega)}{r}\right)^2+\frac{a(\omega)}{r^2},
\end{equation}
associated to $q$ (Friedrichs' Extension) is well defined on a domain containing $L^2(\R^d)$,
therefore the $L^2$-initial value problem
\begin{equation}  \label{eq:schroH}
\begin{cases}
i\partial_t\psi=-iH\psi,
\\
\psi(0)=\psi_0\in L^2(\mathbb R^d),
\end{cases}
\end{equation}
for the wavefunction $\psi=\psi(t,x):\mathbb R\times\mathbb R^d\to\mathbb C$ makes sense.
Here $d\geq2$, and we choose a {\it transversal gauge} for the magnetic vector potential, i.e. we assume 
\begin{equation}  \label{eq:transversality}
{\mathbf{A}}(\omega)\cdot\omega=0 \quad \text{for all }\omega\in {\mathbb{S}}
^{d-1}.
\end{equation}
Notice
that equation \eqref{eq:schroH} is invariant under the scaling
$u_\lambda(x,t):=u(x/\lambda, t/\lambda^2)$, which is the same of the free
Schr\"odinger equation. 

The aim is to understand the role of the spherical operator $L$ associated to $H$,
defined by
\begin{equation}  \label{eq:laplacebeltrami}
L =\big(-i\,\nabla_{\mathbb{S}^{d-1}}+{\mathbf{A}}\big)^2+a(\omega),
\end{equation}
where $\nabla_{\mathbb{S}^{d-1}}$ is the spherical gradient on the unit
sphere $\mathbb{S}^{d-1}$. 
The spectrum of the operator $L$ is formed by a
diverging sequence of real eigenvalues with finite multiplicity
$\mu_0({\mathbf{A}}
,a)\leq\mu_1({\mathbf{A}},a)\leq\cdots\leq\mu_k({\mathbf{A}},a)\leq\cdots$
(see e.g. \cite[Lemma A.5]{FFT}), where each eigenvalue is repeated
according to its multiplicity. Moreover we have that
$\lim_{k\to\infty}\mu_k({\mathbf{A}},a)=+\infty$.  To each $k\geq1$,
we can associate a
$L^{2}\big({\mathbb{S}}^{d-1},{\mathbb{C}}\big)$-normalized
eigenfunction $\varphi_k$ of the operator $L$ on
$\mathbb{S} ^{d-1}$ corresponding to the $k$-th eigenvalue
$\mu_{k}({\mathbf{A}},a)$, i.e.  satisfying
\begin{equation}  \label{eq:angular}
\begin{cases}
L\varphi_{k}=\mu_k({\mathbf{A}},a)\,\varphi_k, & \text{in
}{\mathbb{S}}^{d-1}, \\[3pt]
\int_{{\mathbb{S}}^{d-1}}|\varphi_k|^2\,dS(\theta)=1. &
\end{cases}
\end{equation}
In particular, if $d=2$, $\varphi_k$ are one-variable $2\pi$-periodic functions, i.e. $\varphi_k(0) = \varphi_k(2\pi)$. Since the
eigenvalues $\mu_k({\mathbf{A}},a)$ are repeated according to their 
multiplicity, exactly one eigenfunction $\varphi_k$ corresponds to
each index $k\geq1$. We can choose the functions $\varphi_k$ in such
a way that they form an orthonormal basis of $L^2({\mathbb{S} }^{d-1},{\mathbb{C}})$. We also introduce the numbers
\begin{equation}  \label{eq:alfabeta}
\alpha_k:=\frac{d-2}{2}-\sqrt{\bigg(\frac{d-2}{2}\bigg)^{\!\!2}+\mu_k({\
\mathbf{A}},a)}, \quad \beta_k:=\sqrt{\left(\frac{d-2}{2}\right)^{\!\!2}+
\mu_k({\mathbf{A}},a)},
\end{equation}
so that $\beta_{k}=\frac{d-2}{2}-\alpha_{k}$, for $k=1,2,\dots$.

Under the condition
\begin{equation}  \label{eq:hardycondition}
\mu_0({\mathbf{A}},a)>-\frac{(d-2)^2}{4}
\end{equation}
the quadratic form $q$ in \eqref{eq:quaform} associated to $H$ is positive
definite, and the Friedrichs' extension of $H$ is well defined, with domain
\begin{equation}  \label{eq:domain}
\mathcal{D}:=\left\{ f\in H^1_*({\mathbb{R}}
^d):\ Hf\in L^2({\mathbb{R}}^d\right\},
\end{equation}
where $H^1_*({\mathbb{R}}^d)$ is the completion of $C^{\infty}_{\mathrm{c} }(
{\mathbb{R}}^d\setminus\{0\},{\mathbb{C}})$ with respect to the norm
\begin{equation*}
\|f\|_{H^1_*({\mathbb{R}}^d)}=\bigg(\int_{{\mathbb{R}}^N}\bigg(
|\nabla f(x)|^2+ \frac{|f(x)|^2}{|x|^2}+|f(x)|^2\Big)\bigg) \,dx
\bigg)^{\!\!1/2}.
\end{equation*}
By the Hardy's inequality \eqref{eq:hardy},
$H^1_*({\mathbb{R}}^d)=H^1({\mathbb{R}}^d)$ with equivalent norms if
$d\geq 3$, while $H^1_*({\mathbb{R}}^d)$ is strictly smaller than
$H^1({\mathbb{R}}^d)$ if $d=2$. Furthermore, from condition
\eqref{eq:hardycondition} and \cite[Lemma 2.2]{FFT}, it follows that $H^1_*({\mathbb{R}}^d)$ coincides
with the space obtained by completion of $C^{\infty}_{\mathrm{c}
}({\mathbb{R}}^d\setminus\{0\},{\mathbb{C}})$ with respect to the norm naturally
associated to $H$, i.e.
\begin{equation*}
  q[\psi]+\|\psi\|_2^2.
\end{equation*}

We remark that $H$ could be not essentially
self-adjoint. Indeed, in the case ${\mathbf{A}}\equiv0$, Kalf, Schmincke, Walter, and W\"ust \cite{kwss} and
Simon \cite{simon73}
proved that $H$ is essentially
self-adjoint if and only if $\mu_0(\mathbf{0},a)\geq -\big(\frac{d-2}{2}\big)^{2}+1$ and, consequently, admits a
unique self-adjoint extension (which coincides with the Friedrichs'
extension); otherwise, i.e. if $\mu_0(\mathbf{0},a)<
-\big(\frac{d-2}{2}\big)^{2}+1$, $H$ is not essentially self-adjoint and
admits infinitely many self-adjoint extensions, among which the Friedrichs'
extension is the only one whose domain is included in the domain of
the associated quadratic form (see also \cite[Remark
2.5]{duyckaerts}).

The Friedrichs' extension $H$ naturally
extends to a self adjoint operator on the dual $\mathcal D^\star$ of
$\mathcal{D}$ and the unitary group $e^{-itH}$
extends to a group of isometries on
the dual of $\mathcal{D}$ which will be
still denoted as $e^{-itH}$ (see
\cite{Cazenave}, Section 1.6 for further details). Then for every
$\psi_0\in L^2({  \mathbb{R}}^d)$, 
\begin{equation*}
\psi(t,x):=e^{-itH}\psi_0(x)\in
\mathcal{C}(\mathbb{R};L^2({\mathbb{R}}^d))\cap
  \mathcal C^1({\mathbb{R}};
\mathcal D^\star),
\end{equation*}
is the unique solution to \eqref{eq:schroH}.

Now, by means of \eqref{eq:angular} and \eqref{eq:alfabeta} define the
following kernel:
\begin{equation}  \label{nucleo}
K(x,y)=\sum\limits_{k=-\infty}^{\infty }i^{-\beta _{k}}j_{-\alpha
_{k}}(|x||y|)\varphi _{k}\big(\tfrac{x}{|x|}\big)\overline{\varphi _{k}\big(\tfrac{
y}{|y|}\big)},
\end{equation}
where
\begin{equation*}
j_{\nu }(r):=r^{-\frac{d-2}{2}}J_{\nu +\frac{d-2}{2}}(r)
\end{equation*}
and $J_{\nu }$ denotes the usual Bessel function of the first kind
\begin{equation*}
J_{\nu }(t)=\bigg(\frac{t}{2}\bigg)^{\!\!\nu }\sum\limits_{k=0}^{\infty }
\dfrac{(-1)^{k}}{\Gamma (k+1)\Gamma (k+\nu +1)}\bigg(\frac{t}{2}\bigg) 
^{\!\!2k}.
\end{equation*}
Notice that \eqref{nucleo} reduces to \eqref{eq:jacobi}, in the free case $\mathbf {A}\equiv a\equiv0$.
The first result we mention in this survey is the following representation formula for $e^{-itH}$:
\begin{theorem}[L. Fanelli, V. Felli, M. Fontelos, A. Primo - \cite{FFFP}]\label{thm:1}
Let $d\geq3$, $a\in
L^{\infty }({\mathbb{S}}^{d-1},{\mathbb{R}})$ and ${\ \mathbf{A}}\in C^{1}({
\mathbb{S}}^{d-1},{\mathbb{R}}^{N})$, and assume \eqref{eq:transversality}
and \eqref{eq:hardycondition}. Then, for any $\psi_0\in L^2({\mathbb{R}}^d)$,
\begin{equation}  \label{eq:representation}
e^{-itH}\psi_0(x)=\frac{e^{\frac{i|x|^{2}}{4t}}}{
i(2t)^{{d}/{2}}}\int_{{\mathbb{R}} ^{d}}K\bigg(\frac{x}{\sqrt{2t}},\frac{y}{
\sqrt{2t}}\bigg)e^{i\frac{|y|^{2}}{ 4t}}\psi_{0}(y)\,dy.
\end{equation}
\end{theorem}
As an immediate consequence, we see by \eqref{eq:representation} that the analog to condition \eqref{eq:condiz} gives  for $H$ the complete list of usual time decay estimates \eqref{eq:decayfull}:
\begin{corollary}\label{cor:1}
Let $d\geq3$, $a\in
L^{\infty }({\mathbb{S}}^{d-1},{\mathbb{R}})$ and ${\ \mathbf{A}}\in C^{1}({
\mathbb{S}}^{d-1},{\mathbb{R}}^{N})$, and assume \eqref{eq:transversality}
and \eqref{eq:hardycondition}. If
\begin{equation}\label{eq:condiz2}
\sup_{x,y\in\mathbb R^d}|K(x,y)|<\infty,
\end{equation}
then
\begin{equation}\label{eq:decayfullH}
\left\|e^{-itH}\psi_0(\cdot)\right\|_{L^p(\R^d)}
\leq C|t|^{-d\left(\frac12-\frac1p\right)}\|\psi_0\|_{L^{p'}(\R^d)},
\qquad
\forall t\in\R,
\qquad
\forall p\geq2,
\end{equation}
for some $C>0$ independent on $\psi_0$.
\end{corollary}
In the two last decades, estimates \eqref{eq:decayfullH} were intensively studiedby several authors. The following is an incomplete list of results about this topic \cite{BG, EGS1, EGS2, DF1, DF2, G, GS, PSTZ, RS, S, W1, W2, Y1, Y2, Y3, Y4}. In all these papers, the potentials are sub-critical with respect to the functional scale of the Hardy's inequality \eqref{eq:hardy}: in other words, the critical potentials in \eqref{eq:hamiltonianH} are never considered, and it does not seem that one could handle them by perturbation techniques, which are a common factor of alle the above mentioned papers. Now, formula \eqref{eq:representation} and Corollary \ref{cor:1} give a usual tool to reduce matters to prove time decay, to a spectral analysis problem. This allowed us to prove some new positive results concerning with estimates \eqref{eq:decayfullH}. In 2D, the picture is quite well understood, thanks to the following theorem.
\begin{theorem}[L. Fanelli, V. Felli, M. Fontelos, A. Primo - \cite{FFFP2}]\label{thm:2}
Let $d=2$, $a\in
  W^{1,\infty}(\mathbb{S}^1,\mathbb{R})$, $\mathbf{A}\in
  W^{1,\infty}(\mathbb{S}^1,\mathbb{R}^2)$ satisfying
  \eqref{eq:transversality} and $\mu_1(\A,a)>0$, and
  $H$ be given by
  \eqref{eq:hamiltonianH}. Then, for any $\psi_0\in
  L^2({\mathbb{R}}^d)\cap L^{p'}({ \mathbb{R}}^d)$, 
  \begin{equation}  \label{eq:decayfull2D}
\left\|e^{-itH}\psi_0(\cdot)\right\|_{L^p(\R^2)} 
\leq C|t|^{-2\left(\frac12-\frac1p\right)}\|\psi_0\|_{L^{p'}(\R^2)},
\qquad
\forall t\in\R,
\qquad
\forall p\geq2,
\end{equation}
for some $C>0$ independent on $\psi_0$.
\end{theorem}
Theorem \ref{thm:2} is proved in \cite{FFFP2}. The core consists in proving that \eqref{eq:condiz2} holds, and a crucial role is played by the following Lemma, which gives a quite explicit expansion of eigenvalues and eigenfunctions of $L$, generalizing the results in \cite{Gu}:
\begin{lemma}[L. Fanelli, V. Felli, M. Fontelos, A. Primo - \cite{FFFP2}]\label{lem:spectral}
  Let  $a\in W^{1,\infty}({\mathbb S}^1)$, $\widetilde
a:=\frac1{2\pi}\int_0^{2\pi}a(s)\,ds$, $\mathbf{A}\in W^{1,\infty}({\mathbb
  S}^1)$ such that
\begin{equation}\label{eq:usek}
\widetilde{\mathbf{A}}=\frac1{2\pi}\int_0^{2\pi}A(s)\,ds\not\in
\frac12\Z.
\end{equation}
Then there exist  $k^*,\ell\in\N$ such that $\{\mu_k:\ k>
k^*\}=\{\lambda_j:\ j\in\Z,\ |j|\geq \ell\}$,
\[\sqrt{\lambda_j-\widetilde a}=(\sgn j)
\Big(
\widetilde{\mathbf{A}}-\left\lfloor \widetilde{\mathbf{A}}+\tfrac12\right\rfloor
\Big)
+|j|+O\big(\tfrac1{|j|^3}\big),\quad\text{as }|j|\to+\infty
\]
and
\begin{equation}\label{eq:eigenvalues}
\lambda_j=\widetilde a+
\Big(j+\widetilde{\mathbf{A}}-\big\lfloor \widetilde{\mathbf{A}}+\tfrac12\big\rfloor
\Big)^2+O\big(\tfrac1{j^2}\big),\quad\text{as }|j|\to+\infty.
\end{equation}
Furthermore, for all $j\in\Z$, $|j|\geq\ell$, there exists
 a $L^{2}\big({\mathbb{S}}^{1},{\mathbb{C}}\big)$-normalized
eigenfunction  $\varphi_j$ of the operator $L$ on $\mathbb{S}^{1}$
corresponding to
 the  eigenvalue $\lambda_j$ such that
\begin{align}\label{eq:eigenfunctions}
  \varphi_j(\theta)=
\frac1{\sqrt{2\pi}}e^{-i\big([\widetilde{\mathbf{A}}+1/2]\theta+\int_0^\theta
  A(t)\,dt\big)}\Big(e^{i(\widetilde{\mathbf{A}}+j)\theta}+R_j(\theta)\Big),
\end{align}
where $\|R_j\|_{L^\infty({\mathbb S}^1)}= O\big(\tfrac1{|j|^3}\big)$ as
$|j|\to\infty$. In the above formula $\lfloor\cdot\rfloor$ denotes the floor function
$\lfloor x\rfloor=\max\{k\in\Z:k\leq x\}$.
\end{lemma}
Analogous results to Lemma \ref{lem:spectral} can be proved (and are in part available) in higher dimension $d\geq3$. Nevertheless, the higher dimensional scenario is quite more complicate, and some chaotic behavior of the eigenvalues of $L$ can occur. This makes the generic validity of \eqref{eq:decayfullH} completely unclear in dimension $d\geq3$. In this direction, the only result which is available at the moment is concerned with the 3D-inverse square electric potential, and reads as follows:
\begin{theorem}[L. Fanelli, V. Felli, M. Fontelos, A. Primo - \cite{FFFP}]\label{thm:3}
Let $d=3$, $\mathbf{A}\equiv0$ and $a(\omega)\equiv a\in\mathbb R$, with $a>-\frac14$.
\begin{itemize}
\item[i)] If $a\geq0$, then,  for any $\psi_0\in L^2(\mathbb R^3)\cap L^{p'}(\mathbb R^3)$,
\begin{equation}  \label{eq:decayinverse}
\left\|e^{-itH}\psi_0(\cdot)\right\|_{L^p(\R^2)} 
\leq C|t|^{-3\left(\frac12-\frac1p\right)}\|\psi_0\|_{L^{p'}(\R^2)},
\qquad
\forall t\in\R,
\qquad
\forall p\geq2,
\end{equation}
for some $C>0$ which does not depend on $\psi_0$.

\item[ii)] If $-\frac14<a<0$, let  $\alpha_1$ as in \eqref{eq:alfabeta}, and define
\begin{equation*}
\|\psi\|_{p,\alpha_1}:=\bigg(\int_{{\mathbb{R}}^3}(1+|x|^{-\alpha_1})^{2-p}|\psi(x)|^p\,dx\bigg)^{\!1/p}, \quad
p\geq1.
\end{equation*}
Then the following estimates hold
\begin{equation}  \label{eq:decayinverse2}
\left\|e^{-itH}\psi_0(\cdot)\right\|_{p,\alpha_1}\leq \frac{%
C(1+|t|^{\alpha_0})^{1-\frac{2}{p}}}{|t|^{3\left(\frac12-\frac1p\right)}}%
\|\psi\|_{p',\alpha_0}, \quad p\geq2, \quad \frac1p+\frac1{p^{\prime
}}=1,
\end{equation}
for some constant $C>0$ which does not depend on $\psi_0$.
\end{itemize}
\end{theorem}
\begin{remark}
It is interesting to remark that, in the range $-1/4<a<0$, \eqref{eq:decayinverse} does not hold, while the full set of usual Strichartz estimates hold (see \cite{BPSTZ1, BPSTZ}). This is now clearly understood in terms of formula \eqref{eq:representation}: notice that, if $a=\mu_0<0$, then $\alpha_0>0$ and a negative-index Bessel function appears in the kernel $K$ given by \eqref{nucleo}; since negative-index functions $J_\nu$ are singular at the origin, one cannot either expect the solution \eqref{eq:representation} to be in $L^\infty$. 
\end{remark}
This can be proved as a general fact:
\begin{theorem}[L. Fanelli, V. Felli, M. Fontelos, A. Primo - \cite{FFFP2}]\label{thm:4}
  Let $d\geq3$, $a\in L^{\infty }({\mathbb{S}}^{d-1},{\mathbb{R}})$,
  ${\mathbf{A}}\in C^{1}({\mathbb{S}}^{d-1},{\mathbb{R}}^{d})$, and
  assume \eqref{eq:transversality}, \eqref{eq:hardycondition}, and
  $\mu_0<0$. Then, for almost every $t\in\R$,
  $e^{-itH}(L^1)\not\subseteq L^\infty$; in particular
  $e^{-itH}$ is not a
  bounded operator from $L^1(\mathbb R^d)$ to $L^\infty(\mathbb R^d)$.
\end{theorem}
The above phenomenon can be quantified. To this aim, let us restrict our attention to the case 
\begin{equation*}
  H=-\Delta+\frac{a}{|x|^2},
  \qquad
  x\in\R^3.
\end{equation*}
Let us define
\begin{equation}  \label{eigenvectors}
V_{n,j}(x)= |x|^{-\alpha_j}e^{-\frac{|x|^2}{4}}P_{j,n}\Big(\frac{|x|^2}{2}%
\Big) \psi_j\Big(\frac{x}{|x|}\Big),\quad n,j\in\N,\ j\geq1,
\end{equation}
where $P_{j,n}$ is the polynomial of
degree $n$ given by
\begin{equation*}
P_{j,n}(t)=\sum_{i=0}^n \frac{(-n)_i}{\big(\frac{d}2-\alpha_j\big)_i}\,\frac{%
t^i}{i!},
\end{equation*}
denoting as $(s)_i$, for all $s\in{\mathbb{R}}$, the Pochhammer's symbol 
\begin{equation*}
(s)_i=\prod_{j=0}^{i-1}(s+j),
\qquad
 (s)_0=1.
 \end{equation*}
 Moreover, for all $k> 1$, define
\begin{equation*}
\mathcal U_k=\mathop{\rm span}\left\{V_{n,j}:n\in\N,1\leq j<
  k\right\}\subset L^2(\R^N).
\end{equation*}
The functions $V_{n,j}$ spans $L^2(\R^3)$ (see \cite{FFFP3} for details). Moreover, as initial data for \eqref{eq:schro}, these functions have a quite
explicit evolution: indeed, denoting by $\widetilde V_{n,j}:=V_{n,j}/\|V_{n,j}\|_2$, the following identity holds:
\begin{align}
&
e^{-itH}\widetilde V_{n,j}(x)
=
e^{it\left(-\Delta+\frac{a}{|x|^2}\right)}V_{n,j}(x)
\label{eq:evolution}
\\
&
=    (1+t^2)^{-\frac{d}{4}+\frac{\alpha_j}{2}}\,|x|^{-\alpha_j}
    \frac{e^{\frac{-|x|^2}{4(1+t^2)}}}{\|V_{n,j}\|_{L^{2}(\mathbb{R}^{d})}}
    e^{i\frac{|x|^2 t}{4(1+t^2)}} e^{-i\gamma_{n,j}\arctan t}
    \psi_j\big(\tfrac{x}{|x|}\big)
    P_{n,j}\big(\tfrac{|x|^2}{2(1+t^2)}\big).
    \nonumber
\end{align}
Formula \eqref{eq:evolution} has been proved in \cite{FFFP3}. Clearly, if $a=\mu_0\geq0$, then $\alpha_0\leq0$ and the first function $\widetilde V_{1,0}$ decays polynomially faster than usual, in a weighted space. 
This is reminiscent to question {\bf B} in the Introduction, and gives us the following evolution version of the frequency-dependent Hardy's inequality \eqref{eq:hardyrad3}:
\begin{theorem}[L. Fanelli, V. Felli, M. Fontelos, A. Primo - \cite{FFFP3}]\label{thm:5}
Let $d=3$,  $a=\mu_0\geq 0$, $\alpha_0$ as in \eqref{eq:alfabeta} .
\begin{itemize}
\item[(i)]
There exists $C>0$ such that,
for all $\psi_0\in L^2(\R^3)$ with $|x|^{-\alpha_0}\psi_0\in L^1(\R^3)$,
\begin{equation*}
\left\||x|^{\alpha_0}e^{-itH}\psi_0(\cdot)\right\|_{L^\infty}\leq C
|t|^{-\frac 32+\alpha_0}
\||x|^{-\alpha_0}\psi_0\|_{L^{1}}.
\end{equation*}
\item[(ii)]
  For all
$k\in\N$, $k\geq1$, there exists $C_k>0$ such that,
for all $\psi_0\in \mathcal U_k^\perp$ with $|x|^{-\alpha_k}\psi_0\in L^1(\R^3)$,
\begin{equation*}
\left\||x|^{\alpha_k}e^{-itH}\psi_0(\cdot)\right\|_{L^\infty}\leq C_k
|t|^{-\frac 32+\alpha_k}
\||x|^{-\alpha_k}\psi_0\|_{L^{1}}.
\end{equation*}
\end{itemize}
\end{theorem}
Some analogous results, only concerning with the decay of the first frequency space, had been previously proven in \cite{FGK, GK}.

\par \ \par\ 
To complete the survey, we leave some open questions.
\begin{enumerate}
\item[(i)]
Concerning Theorems \ref{thm:2}, \ref{thm:3}, does any general result hold in dimension $d\geq3$?
\item[(ii)]
In what extent can one perturb the models in \eqref{eq:hamiltonianH}? What is the real role played by the scaling invariance?
\item[(iii)]
The proof of formula \eqref{eq:representation} strongly relies on some pseudoconformal law associated to the free Schr\"odinger flow (Appell transform; see \cite{FFFP}). Is there any analog for other dispersive models, e.g. the wave equation?
\item[(iv)]
One can use formula \eqref{eq:representation} to represent the wave operators
\begin{equation*}
W_{\pm}:=L^2-\lim_{t\to \pm\infty}e^{itH}e^{-itH_0},
\qquad
H_0:=-\Delta.
\end{equation*}
What can one prove about the boundedness of $W_{\pm}$ in $L^p(\R^d)$, in the same style as in \cite{W1, W2, Y1, Y2, Y3, Y4} (at least in 2D, having in mind Theorem \ref{thm:2}.
\item[(v)]
By standard $TT^\star$-arguments, one can obtain some weighted Strichartz estimates by Theorem \ref{thm:5}. Which kind of informations do these estimates give for nonlinear Schr\"odinger equations associated to $H$?
\end{enumerate}

\begin{acknowledgement}
The author sincerely thanks all the organizers of the event {\it Contemporary Trends in the Mathematics of Quantum Mechanics} for the kind invitation, in particular Alessandro Michelangeli for his job. A special acknowledgement is to Prof. Gianfausto Dell'Antonio, for his beautiful scientific contributions, for his teachings, and for his friendship.
\end{acknowledgement}

\end{document}